\documentclass[a4paper]{article}
\usepackage{INTERSPEECH2019}

\usepackage{graphicx}
\usepackage{amssymb,amsmath,bm}
\usepackage{textcomp}

\usepackage{amsmath,graphicx}
\usepackage[font=small]{caption}
\usepackage{tabularx}
\usepackage{threeparttable}
\usepackage{subcaption}
\usepackage{dblfloatfix}
\usepackage{fixltx2e}
\usepackage{multirow}
\usepackage{booktabs}
\usepackage{balance}
\usepackage{amssymb}
\usepackage[compress]{cite}
\graphicspath{{./}}
\DeclareGraphicsExtensions{.pdf,.png,.eps}
\usepackage{epstopdf}

\ninept

\title{{EML} System Description for VoxCeleb Speaker Diarization Challenge 2020 }

\name{Omid Ghahabi, Volker Fischer}
\address{EML Speech Technology GmbH, Berliner Stra{\ss}e 45, 69120 Heidelberg, Germany}
 \email{(omid.ghahabi | volker.fischer)@eml.org}

\begin{document}

  \maketitle
  \begin{abstract}
This technical report describes the EML submission to the first VoxCeleb speaker diarization challenge. 
Although the aim of the challenge has been the offline processing of the signals, the submitted system is basically 
the EML online algorithm which decides about the speaker labels in runtime approximately every 1.2 sec. 
For the first phase of the challenge, only VoxCeleb2 dev dataset was used for training. The results on the provided VoxConverse 
dev set show much better accuracy in terms of both DER and JER compared to the offline baseline provided in the challenge. 
The real-time factor of the whole 
diarization process is about 0.01 using a single CPU machine. 
\end{abstract}
 \vspace{0.2cm} 

\section{Introduction}
\label{sec:intro}
Speaker diarization aims to determine who is speaking when in a multi-speaker conversation. 
The identity and the number of the speakers are typically unknown to the system which makes diarization more difficult 
compared to other common speaker recognition tasks.
Compared to other speaker diarization challenges, the emphasis in the VoxCeleb speaker diarization challenge 
 has been on more challenging conditions found in online videos ``in the wild'', like short rapid speaker exchanges, 
 lots of cross-talks, background degradation by channel noise, music, laughter, and applause \cite{chung2020spot}. 

Although the challenge has been open in terms of training data, we used only VoxCeleb2 dev dataset \cite{chung2018voxceleb2}
which is about 2300~h with 6000 labeled speakers. The voice activity detection (VAD) is an unsupervised version of the method proposed
in \cite{ghahabi_ESSV_VAD_2018} and the speaker diarization is a modified version of the online algorithm proposed in \cite{ghahabi2019speaker}.
More details are given in the following sections.

\section{System Description}
\label{sec:system}
The participating system is a modified version of the online algorithm proposed in \cite{ghahabi2019speaker} which includes three main parts, namely
VAD, speaker embeddings, and the online diarization algorithm itself. 

\subsection{Voice Activity Detection}
\label{subsec:VAD}
An unsupervised version of the VAD algorithm proposed in \cite{ghahabi_ESSV_VAD_2018} is used for this challenge. 
First, a GMM-based Universal Background Model (UBM) is trained. Then given the UBM, a zero-order Baum-Welch statistic vector, referred to as VAD vector, is extracted for 
every 20 frames of background data. Next, VAD vectors are classified into two classes namely speech and non-speech using K-means algorithm. The cluster centroids are kept
as VAD models. In the testing phase, the unknown feature vectors are compared to VAD models via cosine distance.

\subsection{Speaker Embeddings}
\label{subsec:SpeakerEmbeddings}
The speaker embeddings are similar to those we proposed in \cite{ghahabi2019speaker} with two main changes. 
Rather than raw MFCCs, we have transformed them to a more speaker discriminative space considering a context of 
$\left\{-6,-4,-2,0,2,4,6\right\}$ and unsupervised phonetic labels. The network architecture is also changed to 
$\left[3840\; 1200\; 600\; 600\; 300\; 300\; 5000\right]$ rather than having the same hidden layer sizes in \cite{ghahabi2019speaker}.
In summary, the network is a feedforward neural network with VReLU \cite{ghahabi2018restricted, ghahabi2019speaker} 
activation function, low dimensional supervectors as inputs, 5000 speaker classes as outputs, and the 
last hidden layer for speaker embeddings. VReLU allows us to merge all the hidden layers in a single 
transformation matrix from supervectors to speaker embeddings in the testing phase which will reduce the computational cost 
to a great extent in runtime. 

\subsection{Online Diarization Algorithm}
\label{subsec:online-algorithm}
\begin{figure}[t]
\centering
\includegraphics[width=0.43\textwidth]{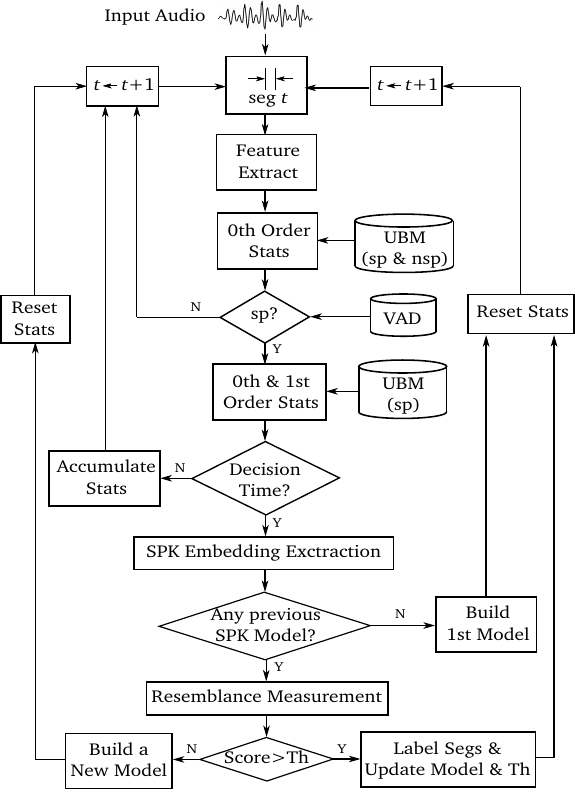}
\caption{EML online speaker diarization algorithm.} 
\label{fig:flowchart}
\end{figure}

Figure~\ref{fig:flowchart} summarizes the proposed online algorithm in \cite{ghahabi2019speaker} in which the input audio signal
is processed segment by segment every 0.2 sec to decide if the current segment is speech (sp) or nonspeech (nsp)
based on the VAD vectors extracted in Section~\ref{subsec:VAD}.
If the short segment is detected as nonspeech, it will be discarded otherwise
the Baum-Welch statistics are computed and accumulated over speech segments until the decision time is reached.
Decision time is based on predefined maximum speech 
or successive nonspeech durations which are set to 2.4 sec and 0.6 sec for this challenge, respectively. 
Accumulated statistics are converted to a low dimensional supervector which is further 
centralized by the 
UBM mean supervector. The resulting supervector is then converted to a lower dimensional speaker embedding.
The speaker embedding is compared with the current speaker models using cosine similarity.
If the model with the highest similarity gives a score higher than the speaker-dependent threshold, 
all the segments corresponding to the accumulated statistics 
are labeled as the selected speaker and the speaker model and its threshold are updated if the scores have been reliable enough. 
Otherwise,
a new model will be created. Before a new model creation, the speech segment is divided into two halves. 
For each half, a speaker vector is created and compared with another. If the two halves are 
similar enough in terms of the speaker identity, the statistics are merged and the new model is created. Otherwise,
each half is assigned to one of the current speaker models. In other words, a new speaker model is created only if two 
halves are similar enough.

\section{Experiments}
\label{sec:experiments}
Diarization Error Rate (DER) is used as the primary performance measurement
in this challenge. DER is defined as the sum of missed speech (MS), false
alarm speech (FA), and speaker misclassification error (speaker
confusion, SC). A forgiveness collar of 0.25 sec is applied
in order to compensate for small inconsistencies in annotation. Overlapping speech is also 
taken into account in evaluation. 

\subsection{Database and Setup}
\label{subsec:Data and Setup}
The VoxConverse dev set is provided for development in the challenge \cite{chung2020spot}. It consists of 216
multi-speaker audios covering 1,218 minutes with 8,268 speaker turns annotated. Some audios have about up to 30\%
overlapping speech from different speakers. More details about the development set can be found in \cite{chung2020spot}.

For unsupervised VAD training, we have used VoxCeleb2 dev set plus 1500~h in house broadcast news data. 
For speaker embedding training, we have used only VoxCeleb2 dev set without any augmentation. After removing some speakers having less than 2 min speech,
5000 speakers are picked for the neural network training.  

Feature vectors are extracted every 10 msec with a 25 msec window. There are 16 dimensional MFCCs along with their deltas for VAD,
and 30 dimensional MFCCs plus their delta for speaker embeddings. VAD features are mean normalized with a 3~sec sliding window. 
UBM for both VAD and speaker vectors are GMMs with 64 Gaussian mixtures each.

\subsection{Results}
\label{subsec:Results}
The preliminary results on the VoxConverse dev set is compared with the challenge baseline system in Table~\ref{table:DER-dev-results}.
In summary, the challenge baseline is basically the same baseline provided in DIHARD 2019 evaluation \cite{ryant2019second} in which 
speech segments are divided into 1.5 sec segments with 50\% overlap. Speaker embeddings are 
x-vectors scored by PLDA. Segments are then clustered using AHC algorithm. More detail about the baseline system can be found in 
\cite{chung2020spot}.

\begin{table}[t!]
\centering
\begin{threeparttable}[b]
  \footnotesize
  \setlength{\tabcolsep}{7.8pt}
 \renewcommand{\arraystretch}{1.1}
 \caption{DER comparison of the EML online system with the challenge baselines on VoxConverse dev set.}\vspace{-0.2cm}
 \label{table:DER-dev-results}
\begin{tabular}{lcccc}
\hline
\textbf{\begin{tabular}[c]{@{}l@{}}\end{tabular}} & \textbf{MS} & \textbf{FA} & \textbf{SC} & \textbf{DER} \\ \cline{1-5} 
Challenge Baseline \cite{chung2020spot}                                                       & 11.1           & 1.4         & 11.3               & 23.8        \\ \hline
Challenge Baseline\tnote{*}~  \cite{chung2020spot}                                                           & 9.3           & 1.3         & 9.7               & 20.2        \\ \hline
EML Online System                                                         & \textbf{3.0}              &   \textbf{0.8}          &   \textbf{9.3}                 & \textbf{13.10}         \\ \hline\vspace{-0.2cm}
\end{tabular}
   \begin{tablenotes}
     \item[*] with speech enhancement
   \end{tablenotes}
\end{threeparttable}
\end{table}


%
%
%

\section{Conclusions}
\label{sec:conclusions}
We have briefly described the EML online speaker diarization system for participating in the VoxCeleb speaker diarization 
challenge 2020. Although the challenge has been based on offline processing of the audio multi-speaker signals and it has been 
an open condition in terms of training data, we participated with the EML online algorithm and using only VoxCeleb2 dev dataset
for training.
The results are significantly better than the challenge baseline systems on the dev set with processing speed of 
0.01$\times RT$ running on a single CPU machine which is considerably much faster than typical offline algorithms. 
However, the results on the evaluation set did not show the same gain compared to the baseline until the deadline for the 
challenge workshop. This could be due to some inappropriate parameter tuning scheme and also a few data used for 
speaker embedding training.

\balance
\bibliographystyle{IEEEtran}

\bibliography{Mybib}


\end{document}